\documentclass[12pt]{article}
\usepackage{amssymb}
\usepackage{amstext}
\usepackage{theorem}
\usepackage{stmaryrd}
\usepackage{cite}
\usepackage{graphicx}
\usepackage{rotating}
\usepackage[normalem]{ulem}

\renewcommand{\sout}[1]{}

\title{
\huge\bf $\,$\\[-4ex]
Coding for Errors and Erasures
in Random Network Coding}

\author{
   {\bf Ralf K\"otter}\thanks{Supported in part by DARPA ITMANET W911NF-07-I-0029.}\\
   Institute for Communications Engineering\\
   TU Munich\\
   D-80333 Munich\\
   {\tt ralf.koetter@tum.de}
\and
   {\bf Frank R. Kschischang}\\
   The Edward S. Rogers Sr.\ Department\\
   of Electrical and Computer Engineering\\
   University of Toronto\\
   {\tt frank@comm.utoronto.ca}}

\date{\small Submitted to {\sl IEEE Transactions on Information Theory}\\Submission Date: March 12, 2007. Revised: March 18, 2008.}

\advance \topmargin by -\headheight
\advance \topmargin by -\headsep
     
\evensidemargin \oddsidemargin
\DeclareMathAlphabet{\mathbfsl}{OT1}{cmr}{bx}{it}

\newcommand{\cut}[1]{}

\renewcommand{\Bbb}{\mathbb}

\newcommand{\F}{\Bbb F}

\newcommand{\Z}{\Bbb{Z}}

\renewcommand{\leq}{\leqslant}

\renewcommand{\geq}{\geqslant}

\newcommand{\highsup}[1]{\raisebox{0.35ex}{\kern 1pt $\scriptstyle {#1}
$}}

\newcommand{\ec}{\end{center}}

\newcommand{\Proof}{\hspace*{10pt}{{\it Proof. }}}

\def\qed{\hskip 3pt \hbox{\vrule width4pt depth2pt height6pt}}

\theoremstyle{plain}
\theorembodyfont{\normalfont\slshape}

\newtheorem{thm}{Theorem}

\newtheorem{lem}[thm]{Lemma}

\newtheorem{cor}[thm]{Corollary}

\newtheorem{defi}{Definition}

\theorembodyfont{\normalfont}

\newtheorem{exam}{Example}

\newtheorem{algo}{Algorithm}

\newtheorem{rem}{Remark}

\outer\def\proclaim #1. #2\par{\medbreak
 \noindent{\bf#1.\enspace}{\sl#2\par}%
 \ifdim\lastskip<\medskipamount \removelastskip\penalty55\medskip\fi}

\makeatletter

\gdef\@punct{.\ \ }  
\def\@sect#1#2#3#4#5#6[#7]#8{%
  \ifnum #2>\c@secnumdepth
     \def\@svsec{}
  \else
     \refstepcounter{#1}\edef\@svsec{%
     \ifnum #2>0{{\csname the#1\endcsname}}.\fi%
    \hskip .5em}
  \fi
  \@tempskipa #5\relax
  \ifdim \@tempskipa>\z@
     \begingroup #6\relax
       \@hangfrom{\hskip #3\relax\@svsec}{\interlinepenalty \@M #8\par}
     \endgroup
     \csname #1mark\endcsname{#7}
     \addcontentsline{toc}{#1}{\ifnum #2>\c@secnumdepth\else
          \protect\numberline{\csname the#1\endcsname}\fi#7}
  \else
     \def\@svsechd{#6\hskip #3\@svsec #8\@punct\csname
#1mark\endcsname{#7}
     \addcontentsline{toc}{#1}{\ifnum #2>\c@secnumdepth \else
          \protect\numberline{\csname the#1\endcsname}\fi#7}}
  \fi
  \@xsect{#5}}

\def\@ssect#1#2#3#4#5{\@tempskipa #3\relax
  \ifdim \@tempskipa>\z@
     \begingroup #4\@hangfrom{\hskip #1}{\interlinepenalty \@M
#5\par}\endgroup
  \else \def\@svsechd{#4\hskip #1\relax #5\@punct}\fi
  \@xsect{#3}}

\makeatother

\newcommand{\gc}[3]{{{#1}\brack{#2}}_{#3}}

\renewcommand{\dim}[1]{\mathop{\mathrm{dim}}({#1})}

\newcommand{\lc}{{\rm lc}}
\newcommand{\lt}{{\rm lt}}
\newcommand{\ev}{{\sf ev}}

\begin{document}
\maketitle

\begin{abstract}
The problem of error-control in random
linear network coding is considered.
A ``noncoherent'' or ``channel oblivious''
model is assumed where neither transmitter nor receiver
is assumed to have knowledge of the channel transfer characteristic.
Motivated by the property that
linear network coding is vector-space preserving,
information transmission is modelled
as the injection into the network of a basis for
a vector space $V$ and the collection by the receiver
of a basis for a vector space $U$.
A metric on the projective geometry associated
with the packet space is introduced,
and it is shown that a minimum
distance decoder for this metric achieves correct
decoding if the dimension of the space $V \cap U$
is sufficiently large.
If the dimension of each codeword is restricted to
a fixed integer, the code forms a subset of a finite-field Grassmannian,
or, equivalently, a subset of the vertices of
the corresponding Grassmann graph.
Sphere-packing and sphere-covering bounds as well as
a generalization of the Singleton bound are provided for
such codes.
Finally, a Reed-Solomon-like code construction,
related to Gabidulin's construction of maximum rank-distance codes,
is described
and a Sudan-style ``list-1'' minimum distance decoding algorithm is 
provided.
\end{abstract}


\clearpage

\section{Introduction}
\label{sec:introduction}

Random network coding \cite{HKMKE03,CWJ03,HMKKESL06}
is a powerful tool for
disseminating information in networks, yet it
is susceptible to packet transmission errors
caused by noise or intentional jamming.
Indeed, in the most naive implementations, a single error
in one received packet
would typically render the entire transmission
useless when the erroneous packet is combined with
other received packets to deduce the transmitted message.
It might also happen that insufficiently
many packets from one generation reach the intended
receivers, so that the problem of deducing the information
cannot be completed.

In this paper we formulate a coding theory in the context of
a ``noncoherent'' or ``channel oblivious''
transmission model for random linear network coding
that captures the effects both of errors, i.e.,
erroneously received packets, and of erasures, i.e.,
insufficiently many received packets.
We are partly motivated by the close analogy
between the $\F_q$-linear channel produced in random linear
network coding
and the ${\mathbb C}$-linear channel produced
in noncoherent multiple-antenna
channels \cite{ZhTs02}, where neither the
transmitter nor the receiver
is assumed to have knowledge of the channel transfer
characteristic.
In contrast
with previous approaches to error control
in random linear network coding, e.g., \cite{CY02,YC06i,YC06ii,Zh08},
the noncoherent transmission strategy taken in this paper
is oblivious to the underlying
network topology and to the particular linear network
coding operations performed at the various network nodes.
Here, information is encoded in the choice at the transmitter
of a \emph{vector space} (not a vector), and
the choice of vector space is conveyed via transmission
of a generating set for the space.

Just as codes defined on the complex Grassmann manifold
play an important role in noncoherent multiple-antenna
channels \cite{ZhTs02},
we find that codes defined in an appropriate
Grassmannian associated with a vector space over
a finite field play an important role here, but
with a different metric associated
with the structure of the corresponding Grassman graph.

The standard, widely advocated approach to random linear
network coding (see, e.g.,
\cite{CWJ03})
involves transmission of packet ``headers'' that
are used to record the particular linear combination
of the components of the message present in each
received packet.
As we will show, this ``uncoded''
transmission may be viewed as a particular code
of subspaces,
but a ``suboptimal'' one, in the
sense that the Grassmannian contains more
spaces of a particular dimension than those obtained
by prepending a header to the transmitted packets.
Indeed, the very notion of a header or local
and global encoding vectors, crucial
in \cite{CWJ03,HMKKESL06,Zh08},
is moot in our context.

A somewhat more closely related approach is
that of \cite{JLKHKM07}, which deals with
reliable communication in networks with so-called ``Byzantine
adversaries,'' who are are assumed to have some ability
to inject packets into the network
and sometimes also to eavesdrop (i.e., read packets transmitted in the network)
\cite{JaLa07}.
It is shown that an optimal communication rate 
(which depends on the adversary's eavesdropping capability)
is achievable with high probability with codes
of sufficiently long block length.
The work of this paper, in contrast, concentrates more
on the possibility of code constructions
with a prescribed deterministic correction capability,
which, however, asymptotically can achieve the 
same rates as would be achieved in the so-called
``omniscient adversary model'' of \cite{JLKHKM07}.

The remainder of this paper is organized as follows.

In Section~\ref{sec:setup}, we introduce 
the ``operator channel'' as a concise and convenient
abstraction of the channel encountered
in random linear network coding, when neither transmitter nor
receiver has knowledge of the channel transfer characteristics.
The input and output alphabet for an operator channel
is the projective geometry (the set of all subspaces)
associated with a given vector space over a finite field $\F_q$.
In Section~\ref{sec:codes}, we define a
metric on this set that is natural and suitable in the context
of random linear network coding.  The transmitter
selects a space $V$ for transmission, indicating this
choice by injection into the network of a set of
packets that generate $V$.
The receiver collects packets that
span some received space $U$.  We show that correct decoding
is possible with a minimum distance decoder if the
dimension of the space $V \cap U$ is sufficiently large,
just as correct decoding in the conventional Hamming
metric is possible if the received vector $u$ agrees
with the transmitted vector $v$ in sufficiently many coordinates.

We will usually confine our attention
to constant-dimension codes, i.e.,
codes in which all codewords have the same dimension.
In this case, the code is a subset of the corresponding Grassmannian,
or, equivalently, a subset of the vertices of the corresponding
Grassmann graph.
Coding in the Grassmann graph has been an active area of research in
combinatorics \cite{Ce84,Chi87,MaZh95,AAK01,SchEtz02},
where the problem has been studied as a
packing problem that arises naturally.
In Section~\ref{sec:bounds},
we derive elementary coding bounds, analogous
to the sphere-packing (Hamming) upper bounds
and the sphere-covering (Gilbert-Varshamov) lower bounds for such codes.
By defining an appropriate notion of puncturing,
we also derive a Singleton bound.  Asymptotic versions
of these bounds are also given.
Some Johnson-type bounds on constant dimension
codes can be found in \cite{XF07}.

A notable application for constant-dimension codes
is their use as so-called linear authentication codes 
introduced
by Wang, Xing and Safavi-Naini \cite{WXSN03}.
There, the properties of codes in the Grassmann
graphs are used to detect tampering with an authenticated message. 
The authors of \cite{WXSN03} describe a construction
by which constant-dimension codes that are suitable
for linear authentication can be obtained from
rank-distance codes, in particular from
the maximum rank-distance codes of Gabidulin \cite{Ga85}.
In Section~\ref{sec:RS} we revisit this construction
in the context of the coding metric defined in this paper,
and we show that these codes achieve the Singleton bound asymptotically.
The main result of Section~\ref{sec:RS} is the development
of an efficient Sudan-style ``list-1'' minimum distance decoding
algorithm for these codes.
A related polynomial-reconstruction-based decoder for Gabidulin
codes has been described by Loidreau \cite{Lo06}.
The connection between rank-metric codes and a generalized
rank-metric decoding problem induced by random linear network
coding is explored in \cite{SiKsKo07}.

\section{Operator Channels}
\label{sec:setup}

We begin by formulating our problem for the case of a single unicast,
i.e., communication between a single transmitter and a single receiver.
Generalization to multicasting is straightforward.

To capture the essence of random linear network coding,
recall \cite{CWJ03,HMKKESL06}
that communication between transmitter and receiver occurs in
a series of rounds or ``generations;''  during each generation,
the transmitter injects a number of fixed-length
packets into the network, each of which may be
regarded as a row vector of length $N$ over a finite field $\F_q$.
These packets propagate through the network, possibly
passing through a number of intermediate nodes between
transmitter and receiver.
Whenever an intermediate node has an opportunity to send a packet, it
creates a random $\F_q$-linear combination of the packets it has
available and transmits this random combination.  Finally, the receiver
collects such
randomly generated packets and tries to infer the set of packets
injected into the network.  There is \emph{no} assumption here
that the network operates synchronously or without delay or that
the network is acyclic.

The set of successful packet transmissions in a
generation induces a directed multigraph with the same vertex set as
the network, in which edges denote successful
packet transmissions.
The rate of information transmission (packets per generation)
between the transmitter
and the receiver is upper-bounded by the min-cut between these
nodes, i.e., by the minimum number of edge deletions in
the graph that would cause the separation of the transmitter
and the receiver.
It is known that random linear network coding in
$\F_q$ is able to achieve a transmission rate that
achieves the min-cut rate with probability approaching one
as $q \rightarrow \infty$ \cite{HMKKESL06}.

Let $\{p_1,p_2,\ldots,p_{M}\}$, $p_i\in \F_q^N$ denote the set of
injected vectors.  In the error-free case, the receiver collects
packets $y_j$, $j=1,2,\ldots,L$ where each $y_j$ is formed as
$y_j=\sum_{i=1}^M h_{j,i} p_i$ with unknown, randomly chosen
coefficients $h_{j,i}\in \F_q$.

We note that \emph{a priori} $L$ is not
fixed and the receiver would normally collect as many packets as possible.
However, as noted above, properties of the network such as the
min-cut between the transmitter and the receiver may influence the joint
distribution of the $h_{i,j}$ and, at some point, there will be no
benefit from collecting further redundant information.

If we choose to consider the injection of $T$ erroneous packets,
this model is enlarged
to include error packets $e_t$, $t = 1, \ldots, T$ to give
\[
y_j=\sum_{i=1}^M h_{j,i} p_i+\sum_{t=1}^T g_{j,t} e_t,
\]
where again $g_{j,t}\in \F_q$ are unknown
random coefficients.  Note that since these erroneous packets may
be injected anywhere within the network, they may
cause widespread error propagation. In particular, if $g_{j,1} \neq 0$
for all $j$, even a
single error packet $e_1$ has the potential to corrupt each and every received
packet.

In matrix form,  the transmission model may be written as
\begin{equation}
y=Hp+Ge,
\label{eq:basic_setup}
\end{equation}
where $H$ and $G$ are random $L\times M$ and $L\times T$ matrices,
respectively,
$p$ is the $M \times N$ matrix whose rows are the transmitted vectors,
$y$ is the $L \times N$ matrix whose rows are the received vectors,
and $e$ is the $T \times N$ matrix whose rows are the error vectors.

The network topology will certainly impose some structure on
the matrices $H$ and $G$.
For example, $H$ may be rank-deficient if the min-cut between transmitter
and receiver is not large enough to support the
transmission of $M$ independent
packets during the lifetime of one generation\footnote{This
statement can be made precise once the precise protocol for transmission
of a generation has been fixed.  However, for the purpose of this paper it
is sufficient to summarily model ``rank deficiency'' as one potential
cause of errors.}.
While the possibility may exist to exploit 
the structure of the network, in
the strategy adopted in this paper
we do not take any possibly finer structure of the
matrix $H$ into account.  Indeed, any such fine structure can be
effectively obliterated by randomization
at the source, i.e., if, rather than injecting packets $p_i$ into
the network, the transmitter were instead
to inject random linear combinations of the $p_i$.
  
At this point, since $H$ is random, we may ask what
property of the injected sequence of packets remains
invariant in the channel described by (\ref{eq:basic_setup}),
even in the absence of noise ($e=0$)?  Since $H$ is a
random matrix, all that is fixed by the product $Hp$
is the row space of $p$.
Indeed, as far as the receiver is concerned,
any of the possible generating sets for this space are equivalent.
We are led, therefore, to consider information transmission
not via the choice of $p$, but rather by the choice of
the vector space spanned by
the rows of $p$.
This simple observation is at the heart of the channel
models and transmission strategies considered in this paper.
Indeed, with regard to the vector space selected by the transmitter,
the only deleterious effect
that a multiplication with
$H$ may have is that
$Hp$ may have smaller rank than $p$, due to, e.g., an insufficient min-cut
or packet erasures, in which case $Hp$ generates a subspace
of the row space of $p$.

Let $W$ be a fixed $N$-dimensional vector space over $\F_q$.
All transmitted and received packets will be vectors of $W$; however,
we will describe a transmission model in terms
of subspaces of $W$ spanned by these packets.
Let $\mathcal{P}(W)$ denote the set of all
subspaces of $W$, an object often called the projective
geometry of $W$.
The dimension of an element $V \in \mathcal{P}(W)$
is denoted as $\dim{V}$.
The sum of two subspaces $U$ and $V$ of $W$
is $U + V = \{ u + v : u \in U,~v\in V \}$.  Equivalently,
$U + V$ is the smallest subspace of $W$ containing both $U$ and $V$.
If $U \cap V = \{0\}$, i.e., if $U$ and $V$ have
trivial intersection, then the sum $U + V$ is
a direct sum, denoted as $U \oplus V$.
Clearly $\dim{U \oplus V} = \dim{U} + \dim{V}$.
For any subspaces $U$ and $V$ we have
$V = (U \cap V) \oplus V'$ for some subspace $V'$
isomorphic to the quotient space $V/(U \cap V)$.
In this case, $U + V = U + ((U \cap V) \oplus V') = U \oplus V'$.

For integer $k \geq 0$,
we define a stochastic operator $\mathcal{H}_k$,
called an ``erasure operator,'' that operates on
the subspaces of $W$.  If $\dim{V}>k$, then
$\mathcal{H}_k(V)$ returns a randomly chosen $k$-dimensional
subspace of $V$;  otherwise, $\mathcal{H}_k(V)$ returns $V$.
For the purposes of this paper, the distribution of
$\mathcal{H}_k(V)$ is unimportant;  for example, it could
be chosen to be uniform.
Given two subspaces $U$ and $V$ of $W$, it is always
possible to realize $U$ as
$U = \mathcal{H}_k(V) \oplus E$
for some subspace $E$ of $W$, assuming that
that $k = \dim{U \cap V}$ and that
$\mathcal{H}_k(U)$ realizes $U \cap V$.

We define the following ``operator channel''
as a concise transmission model for network coding.

\begin{defi}\label{def:operator_channel}
An operator channel $C$ associated with
the ambient space $W$ is a channel with input and
output alphabet $\mathcal{P}(W)$.
As described above,
the channel input $V$
and channel output $U$ can always be related as
\begin{equation}
U = \mathcal{H}_k(V) \oplus E,
  \label{eqn:operatorchannel}
\end{equation}
where $k = \dim{U \cap V}$ and $E$ is an error space.
In transforming $V$ to $U$, we
say that the operator channel commits
$\rho = \dim{V}-k$ \emph{erasures}
and $t = \dim{E}$ \emph{errors}.
\end{defi}

Note that we have chosen to model the error space
$E$ as intersecting trivially with the transmitted
subspace $V$, and thus the choice of $E$ is not independent
of $V$.
However, if we were to model the received space as
$U  = {\cal H}_k(V) + E$ for an arbitrary error space $E$,
then, since $E$ always decomposes for some space $E'$ as
$E = (E \cap V) \oplus E'$, we would get
$U = {\cal H}_k(V) + (E \cap V) \oplus E' = {\cal H}_{k'}(V) \oplus E'$
for some $k' \geq k$.
In other words, components of an
error space $E$ that intersect with the transmitted
space $V$ would only be helpful, possibly
decreasing the number of erasures seen by the receiver.

In summary, an operator channel takes in a vector space
and puts out another vector space, possibly
with erasures (deletion of vectors from the
transmitted space) or errors (addition of vectors to
the transmitted space).

This definition of an operator channel makes a very clear connection
between network coding and classical information theory. Indeed,
an operator channel can be seen as a standard discrete memoryless
channel with input and output 
alphabet $\mathcal{P}(W)$. By imposing a channel law, i.e., transition
probabilities between spaces, it would (at least conceptually)
be straightforward to compute capacity,
error exponents, etc. Indeed, only slight
extensions would be necessary concerning
the ergodic behavior of the channel.
For the present paper we constrain our attention to the question
of constructing good codes 
in $\mathcal{P}(W)$, which is an essentially combinatorial problem.
The codes we construct may be regarded
as ``one-shot'' codes, i.e., codes of length one, since
the transmission of a codeword will induce exactly
one use of the operator channel.

\section{Coding for Operator Channels}
\label{sec:codes}

Definition~\ref{def:operator_channel}
concisely captures the effect of random linear network
coding in the presence of networks with erasures, varying min-cuts
and/or erroneous packets.  Indeed, we will show how to construct codes
for this channel that correct combinations of errors and erasures.
Before we give such a construction we need to define a
suitable metric.

\subsection{A Metric on $\mathcal{P}(W)$}

Let $\Z_+$ denote the set of non-negative integers.
We define a function $d: \mathcal{P}(W) \times \mathcal{P}(W) \rightarrow \Z_+$
by
\begin{equation}
d(A,B) :=\dim{A + B} -\dim{A\cap B}.
\label{eqn:distance}
\end{equation}
Since $\dim{A + B} = \dim{A} + \dim{B} - \dim{A \cap B}$,
we may also write
\begin{eqnarray*}
d(A,B) & = & \dim{A} + \dim{B} - 2 \dim{A \cap B}\\
       & = & 2\dim{A + B} - \dim{A} - \dim{B}.
\end{eqnarray*}

The following lemma is a cornerstone for code design
for the operator channel of Definition~\ref{def:operator_channel}.
\begin{lem}
The function
\[
d(A,B) := \dim{A + B} - \dim{A\cap B}
\]
is a metric for the space $\mathcal{P}(W)$.
\end{lem}

\Proof We need to check that for all subspaces $A, B, X \in \mathcal{P}(W)$
we have:
i) $d(A,B)\geq 0$ with equality if and only if $A=B$,
ii) $d(A,B)=d(B,A)$,
and iii) $d(A,B)\leq d(A,X)+d(X,B)$.
The first two conditions are clearly true and so we focus on the third
condition, the triangle inequality.
%
We have
\begin{eqnarray*}
\frac{1}{2}\left(d(A,B) - d(A,X) - d(X,B)\right) & = &
 \dim{A\cap X} + \dim{B \cap X} - \dim{X} - \dim{A \cap B}\\
& = & \underbrace{\dim{A\cap X + B \cap X} - \dim{X}}_{\leq 0}\\
& & + \underbrace{\dim{A\cap B \cap X} - \dim{A \cap B}}_{\leq 0} \\
& \leq & 0,
\end{eqnarray*}
where the first inequality follows from
the property that $A \cap X + B \cap X \subseteq X$
and the second inequality follows from the
property that $A \cap B \cap X \subseteq A \cap B$.
\qed

\emph{Remark:}  The fact that $d(A,B)$ is a metric also
follows from the fact that this quantity represents the
distance of a geodesic between $A$ and $B$ in the undirected
Hasse graph representing the lattice of subspaces of $W$
\cite{Ce84} partially ordered by inclusion, i.e.,
where $X \preceq Y$ if and only if $X$ is a subspace of $Y$.
In this graph, the vertices correspond to the elements of $\mathcal{P}(W)$
and an edge joins a subspace $X$ with a subspace $Y$
if and only if $| \dim{X} - \dim{Y} | = 1$ and either
$X \subset Y$ or $Y \subset X$.  Just as the hypercube provides
the appropriate setting for coding in the Hamming metric,
the undirected Hasse graph represents the appropriate setting
for coding in the context considered here.\qed

If a basis for $W$ is fixed, then since $W$ is an
$N$-dimensional vector space over $\F_q$, the elements of $W$
may be represented by $N$-tuples of $\F_q$-valued
coordinates with respect to this basis.  We take
the usual inner product between vectors $u=(u_1,\ldots,u_N)$ and
$v =(v_1,\ldots,v_N)$
as $(u,v) = \sum_{i=1}^N u_i v_i$.  If $U$ is a $k$-dimensional
subspace of $W$, then the orthogonal subspace
\[
U^\perp = \{ v \in W: (u,v)=0 \mbox{ for all $u \in U$} \}
\]
is a space of dimension $N-k$.  It is well known
that for any subspaces $U$ and $V$ of $W$ that
$(U^\perp)^\perp = U$,
\[
(U + V)^\perp = U^\perp \cap V^\perp, \mbox{ and }
(U \cap V)^\perp = U^\perp + V^\perp.
\]
It follows that
\begin{eqnarray}
d(U^\perp,V^\perp) & = &
\dim{U^\perp + V^\perp} - \dim{U^\perp \cap V^\perp} \nonumber \\
& = & 
\dim{(U \cap V)^\perp} - \dim{(U + V)^\perp} \nonumber \\
& = &
(N-\dim{U \cap V}) - (N- \dim{U+V}) \nonumber \\
& = & \dim{U+V} - \dim{U \cap V}\nonumber \\
& = & d(U,V) \label{eqn:complement}.
\end{eqnarray}
Thus the distance between subspaces $U$ and $V$ is perfectly
mirrored by the distance between the orthogonal
subspaces $U^\perp$ and $V^\perp$.

\subsection{Codes}

Let $W$ be an $N$-dimensional vector space over $\F_q$.
A \emph{code} for an operator channel with ambient space $W$
is simply a nonempty subset of
$\mathcal{P}(W)$, i.e., a nonempty collection of
subspaces of $W$.

The size of a code $\mathcal{C}$ is denoted by $|\mathcal{C}|$.
The minimum distance of $\mathcal{C}$ is denoted by
\[
D(\mathcal{C}) := \min_{X,Y\in \mathcal{C} :X\neq Y} d(X,Y).
\]
The maximum dimension of the codewords of $\mathcal{C}$
is denoted by
\[
\ell(\mathcal{C}) := \max_{X \in \mathcal{C}}\, \dim{X}.
\]
If the dimension of each codeword of $\mathcal{C}$ is
the same, then $\mathcal{C}$ is said to be a \emph{constant-dimension} code.

In analogy with the $(n,k,d)$ triple that
describes the parameters of a classical linear error
correcting code of length $n$, dimension $k$ and minimum Hamming distance $d$,
a code $\mathcal{C}$ for an operator channel
with an $N$-dimensional ambient space over $\F_q$ is said to be
of type $[N, \ell(\mathcal{C}),\log_q|\mathcal{C}|,D(\mathcal{C})]$.

The \emph{complementary code}
corresponding to a code $\mathcal{C}$
is the code $\mathcal{C}^\perp = \{ U^\perp: U \in \mathcal{C} \}$
obtained from the orthogonal subspaces of the codewords of $\mathcal{C}$.
In view of (\ref{eqn:complement}),
we have $D(\mathcal{C}^\perp) = D(\mathcal{C})$.
If $\mathcal{C}$ is a constant-dimension code of
type $[N,\ell,M,D]$,
then $\mathcal{C}^\perp$ is a constant-dimension code
of type $[N,N-\ell,M,D]$.

Before we study bounds and constructions of codes in $\mathcal{P}(W)$,
we need a proper definition of rate. Let $C \subset \mathcal{P}(W)$
be a code of type $[N,\ell,\log_q|\mathcal{C}|,D]$.  To transmit
a space $V \in \mathcal{C}$ would require the transmitter
to inject up to $\ell(\mathcal{C})$
(basis) vectors from $V$ into the network, corresponding to the
transmission of $N\ell$ $q$-ary symbols.
This motivates the
following definition.

\begin{defi}\label{def:code_params}
Let $\mathcal{C}$ be a code of
type $[N, \ell, \log_q(|\mathcal{C}|), D]$.
The normalized weight $\lambda$,
the rate $R$, and
the normalized minimum distance $\delta$ of $\mathcal{C}$ are defined as
  \[
  \lambda=\frac{\ell}{N},~
  R=\frac{\log_q(|\mathcal{C}|)}{N\ell}~\mbox{and}~
  \delta=\frac{D}{2\ell}.\quad\quad\qed
  \]
\end{defi}

The parameters $\lambda$, $R$ and $\delta$ are quite natural. The
normalized weight $\lambda$ takes the role of the energy of a
spherical code in Euclidean space, or
the equivalent weight parameter for constant weight codes. As such
$\lambda$ is naturally limited to the range $[0,1]$.
For constant-dimension codes, just as in the case of
constant-weight codes, the interesting range can actually be limited to
$[0,\frac{1}{2}]$ as any code $\mathcal{C}$ with $\ell > N/2$
corresponds to the complementary code $\mathcal{C}^\perp$
with $\ell < N/2$ and having the identical distance properties.
The definition of $\delta$ gives a natural range of
$[0,1]$. Indeed, a normalized distance of $1$
could only be obtained by spaces having trivial intersection.
The rate $R$ of a code is restricted to the range $[0,1]$, with a rate of $1$
only being approachable for $\lambda \rightarrow 0$.

The fundamental code construction problem for the operator channel
of Definition \ref{def:operator_channel} thus becomes the determination of
achievable tuples $[\lambda,R,\delta]$ as the dimension of ambient
space $N$ becomes arbitrarily large. We note that this setup
may lack physical reality since it assumes that the network can
operate with arbitrarily long packets; thus we will try to express our
results for finite length $N$ whenever possible.
Furthermore, as noted above, the codes we consider here are ``one-shot'' codes
that induce just a single use of the operator channel.
In situations where the channel characteristics (such
as the network min-cut) are time-varying, it may be interesting
and useful to define codes that induce $n$ uses of the operator
channel, considering
what performance is attainable as $n \rightarrow \infty$.
We will not pursue this direction in this paper.

\subsection{Error and Erasure Correction}

A minimum distance decoder for a code $\mathcal{C}$ is one that
takes the output $U$ of an operator channel and returns a
nearest codeword $V \in \mathcal{C}$, i.e., a codeword $V \in \mathcal{C}$
satisfying, for all $V' \in \mathcal{C}$,  $d(U,V) \leq  d(U,V')$.

The importance of the minimum distance $D(\mathcal{C})$ for a code
$\mathcal{C} \subset \mathcal{P}(W)$ is given in the following theorem,
which provides the combined error-and-erasure-correction capability
of $\mathcal{C}$ under minimum distance decoding.
Define $(x)_+$ as $(x)_+ :=\max\{0,x\}$.

\begin{thm}\label{thm:error_correction}
Assume we use a code $\mathcal{C}$ for transmission over an operator
channel.  Let $V \in \mathcal{C}$ be transmitted, and let
\[
U=\mathcal{H}_k(V)\oplus E
\]
be received, where $\dim{E} = t$.  Let $\rho = (\ell(\mathcal{C})-k)_+$
denote the maximum number of erasures induced by the channel.
If
\begin{equation}
2(t + \rho) < D(\mathcal{C}),
\label{eqn:ecorrection}
\end{equation}
then a minimum distance decoder for $\mathcal{C}$ will produce the
transmitted space $V$ from the received space $U$.
\end{thm}

\Proof Let $V' = {\cal H}_k(V)$.  From the triangle inequality
we have $d(V,U) \leq d(V,V') + d(V',U) \leq \rho + t$.
If $T \neq V$ is any other codeword in $\mathcal{C}$,
then $D(\mathcal{C}) \leq d(V,T) \leq d(V,U) + d(U,T)$,
from which it follows that $d(U,T) \geq D(\mathcal{C}) - d(V,U) \geq
D(\mathcal{C})-(\rho + t)$.  Provided that 
the inequality (\ref{eqn:ecorrection}) holds,
then $d(U,T) > d(U,V)$ and hence a minimum distance decoder
must produce $V$.\qed


Not surprisingly, given the symmetry in this setup
between erasures
(deletion of dimensions due to, e.g., an
insufficient min-cut in the network or an unfortunate
choice of coefficients in the random linear network code) and errors
(insertion of dimensions due to errors or deliberate malfeasance),
erasures and errors are equally costly to the decoder.
This stands in apparent contrast with traditional error correction
(where erasures cost less than errors);
however, this difference is merely an accident of terminology.
A perhaps more closely related classical concept would be that
of ``insertions'' and ``deletions''.

If we can be sure that the projection operation is moot
(expressed by choosing operator $\mathcal{H}_{\dim{W}}$
which operates as an identity on each subspace of $W$)
or that the network produces no errors (expressed
by choosing the error space $E = \{ 0 \}$),
we get the following corollary.

\begin{cor}\label{cor:normal}
Assume we use a code $\mathcal{C}$ for transmission over an operator
channel where $V \in \mathcal{C}$ is transmitted.
If 
\[
U=\mathcal{H}_{\dim{W}}(V)\oplus E = V \oplus E
\]
is received,
and if $2t < D(\mathcal{C})$ where $\dim{E} = t$,
then a minimum distance decoder for $\mathcal{C}$ will produce $V$.
Symmetrically, if
\[
U=\mathcal{H}_{k}(V)\oplus \{ 0 \} = {\cal H}_k(V)
\]
is received, and if $2\rho < D(\mathcal{C})$ where
$\rho = (\ell(\mathcal{C})-k)_+$,
then a minimum distance decoder for $\mathcal{C}$ will produce $V$.
\end{cor}

In other words,
the first part of the corollary states that in the absence of erasures
a minimum distance decoder uniquely
corrects errors up to dimension
\[
t \leq \lfloor \frac{D(\mathcal{C})-1)}{2} \rfloor,
\]
precisely in parallel to the standard
error-correction situation.

\subsection{Constant-Dimension Codes}

In the context of network coding,
it is natural to consider codes in which
each codeword has the same dimension,
as knowledge of the codeword dimension
can be exploited by the decoder
to initiate decoding.
Constant-dimension codes are
analogous to constant-weight codes in Hamming
space (in which every codeword has constant
Hamming weight) or to spherical codes
in Euclidean space (in which every codeword
has constant energy).

As noted above, when considering constant-dimension codes
we may restrict ourselves to codes
of type $[N,\ell,M,D]$ with $\ell \leq N-\ell$, since
a code of type
$[N,\ell,M,D]$ with $\ell > N-\ell$ may be replaced
with its complementary code $C^\perp$ while maintaining
all distance properties (therefore maintaining
all error- and erasure-correcting capability).

Constant-dimension codes
are naturally described as particular vertices
of a so-called
\emph{Grassmann graph}, also called a $q$-Johnson scheme, where the
latter name emphasizes that these objects constitute association
schemes. A formal definition is given as follows.

\begin{defi}\label{def:grass}
Denote by $\mathcal{P}(W,\ell)$ the set of all subspaces of $W$
of dimension $\ell$.   This object is known as a Grassmannian.
The Grassmann graph $G_{W,\ell}$
has vertex set $\mathcal{P}(W,\ell)$ with
an edge joining vertices $U$ and $V$ if and only if $d(U,V)=2$.\qed
\end{defi}

\emph{Remark:}
  It is well known that $G_{W,\ell}$ is distance regular
  \cite{BCN89} and an
  association scheme with relations given by the distance between 
  spaces. As such, practically all techniques for bounds in the
  Hamming association scheme apply. In particular, sphere-packing and
  sphere-covering concepts have a natural equivalent formulation. We explore
  these directions in Section \ref{sec:bounds}.
  We also note that the distance
  between two spaces $U,V$ in $\mathcal{P}(W,\ell)$ introduced 
  in (\ref{eqn:distance}) is, like in the case of constant-weight
  codes in the Hamming metric, an even number
  equal to twice the graph
  distance in the Grassmann
  graph.\footnote{Defining
a distance as half of $d(U,V)$ would give non-integer
values for packings in $\mathcal{P}(W)$.}
As noted Section~\ref{sec:introduction},
codes in the Grassmann graph
have been considered previously
in \cite{Ce84,Chi87,MaZh95,AAK01,SchEtz02,WXSN03}.
  \qed

When modelling the operation of random
linear network coding by the operator
channel of Definition \ref{def:operator_channel}, there is no further
need to specify the precise
network protocol.
In particular, we assume the receiver knows that a codeword $V$ from
$\mathcal{P}(W,\ell)$ was transmitted. In this situation, a receiver
could choose to collect packets until the collected packets,
interpreted as vectors, span an $\ell$-dimensional space. This
situation would correspond to an operator channel of type
$U=\mathcal{H}_{\ell-t(E)}(V)\oplus E$, corresponding to $t(E)$
erasures and $t(E)$ errors. According to Theorem
\ref{thm:error_correction} we can thus correct up to an error dimension
$\lfloor \frac{D(\mathcal{C})-1}{4}\rfloor$. To some extent this
additional factor of two reflects the choice of the distance measure
as being twice the graph distance in $G_{W,\ell}$.
Note that this
situation also would arise if the errors originated in network-coded
transmissions through the min-cut edges in the graph. If the errors do
not affect the min cut but may have arisen anywhere else in the
network, a receiver can choose to collect packets until an $\ell+t(E)$
dimensional space $V\oplus E$ has been recovered.\footnote{Not
  knowing the effective dimension of $E$, i.e.,
  the dimension of $E/(E\cap V)$,
  in practice the receiver would just collect as many packets as
  possible and attempt to reconstruct the corresponding space.} In this case
the error correction capability would increase to an error dimension
$\lfloor \frac{D(\mathcal{C})-1}{2}\rfloor$. We do not study the
implications of this observation further in this paper, since the
coding-theoretic goal of constructing good codes in $\mathcal{P}(W)$ is
not affected by this. Nevertheless, we point out that a properly
designed protocol can (and should) take advantage of these differences.

\subsection{Examples of Codes}

We conclude this section with two examples of codes in $\mathcal{P}(W,\ell)$ .
\begin{exam}\label{exam:header}
  Let $W$ be the vector space of 
  $N$-tuples over $\F_q$.
  Consider the set
  $\mathcal{C}\subset \mathcal{P}(W,\ell)$ of spaces $U_i$,
  $i=1,2,\ldots,|\mathcal{C}|$ with generator matrices
  $G(U_i)=(I|A_i)$ where $I$ is an $\ell \times \ell$ identity
  matrix and the $A_i$ are all possible
  $\ell \times (N-\ell)$
  matrices over $\F_q$.
  It is easy to see that all
  $G(U_i)$ generate different spaces, intersecting in subspaces of
  dimension at most $\ell-1$ and that, hence, the minimum distance of
  the code is $2\ell-2(\ell-1)=2$. The code is a constant-dimension
  code of type $[N,\ell, \ell(N-\ell),2]$ with normalized weight
  $\lambda = \ell/N$, rate
  $R=1-\lambda$ and normalized distance $\delta=\frac{1}{\lambda N}$.\qed
\end{exam}

The first example corresponds to a trivial code that offers no error
protection at all. While this code has been advocated widely for
random linear network coding it is by no means the optimal
code for a given distance $D=2$, as can be seen in the following example.

\begin{exam}\label{exam:Grass2}
  Again let $W$ be the space of vectors of length $N$. We now choose
  the code $\mathcal{C}'=\mathcal{P}(W,\ell)$, which yields a
  constant-dimension code of type
  $[N,\ell, \log_q|\mathcal{P}(W,\ell)|,2]$ which is clearly larger than the
  code $\mathcal{C}$ of Example \ref{exam:header}.  As
  explained in Section~\ref{sec:bounds},
  $|\mathcal{P}(W,\ell)|$ is equal to the
  Gaussian coefficient $\gc{N}{\ell}{q}$.
\end{exam}

We note that $\mathcal{C}''$, defined as
$\mathcal{C}''=\bigcup_{i=1}^{\ell} \mathcal{P}(W,i)$ (no longer
a constant-dimension code) is obviously
an even bigger code (albeit with minimum distance $D=1$)
that can be used for random linear network coding while
not using more network resources than $\mathcal{C}$ and $\mathcal{C}'$.
However, in contrast to $\mathcal{C}'$, the receiver must be able to
determine when the transmission of the code space is complete. This
information is implicit in $\mathcal{C}'$ and $\mathcal{C}$ since the
dimension of the transmitted space is fixed beforehand.\qed

In the next section we provide a few standard bounds for codes in our setup.

\section{Bounds on Codes}
\label{sec:bounds}

\subsection{Preliminaries}

We will be interested in constructing constant-dimension codes.
We start this section by introducing some notation that will be relevant for
packings in $\mathcal{P}(W,\ell)$ where $W$ is an $N$-dimensional
vector space over $\F_q$.

%
%

The $q$-ary \emph{Gaussian coefficient}, the $q$-analogue of
the binomial coefficient, is defined, for non-negative
integers $\ell$ and $n$ with $\ell \leq n$, by
\[
\gc{n}{\ell}{q} := 
\frac{(q^n-1)(q^{n-1}-1)\cdots(q^{n-\ell+1}-1)}{(q^\ell-1)(q^{\ell-1}-1)\cdots(q-1)}
= \prod_{i=0}^{\ell-1} \frac{q^{n-i}-1}{q^{\ell-i}-1},
\]
where the empty product obtained when $\ell=0$ is interpreted as $1$.

As is well known (see, e.g., \cite[Ch.~24]{VLWi01}),
the Gaussian coefficient
$\gc{n}{\ell}{q}$ gives the number of distinct $\ell$-dimensional
subspaces of an $n$-dimensional vector space over $\F_q$.

For $q>1$ the asymptotic behavior of $\gc{n}{\ell}{q}$ is
given given by the following lemma.
\begin{lem}
The Gaussian coefficient $\gc{n}{\ell}{q}$ satisfies
\[1<q^{-\ell(n-\ell)}\gc{n}{\ell}{q}< 4
\]
for $0<\ell<n$, so that we may write
$\gc{n}{\lambda n}{q} = \Theta\left(q^{ n^2\lambda(1-\lambda)}\right)$,
$0 < \lambda < 1$.
\end{lem}

\Proof The quantity $q^{\ell(n-\ell)}$ may be interpreted
as the number of $\ell$-dimensional subspaces 
of $\F_q^n$ that occur as the row space
of a matrix of the form $[I|A]$, where $I$ is an $\ell\times\ell$
identity matrix and $A$ is an arbitrary $\ell \times (n-\ell)$
matrix over $\F_q$.  (This is the number of codewords
in the code $\mathcal{C}$ of Example~\ref{exam:header} with
$N=n$.)  Since  $\ell >0$, this set does not contain
all $\ell$-dimensional subspaces of $\F_q^n$ and
hence the left hand inequality results.
For the right hand inequality we observe that $\gc{n}{\ell}{q}$ may be written as 
\begin{eqnarray*}
\gc{n}{\ell}{q}&=&q^{\ell(n-\ell)}\frac{(1-q^{-n})(1-q^{-n+1})\ldots(1-q^{-n+\ell-1})}{(1-q^{-\ell})(1-q^{-\ell+1})\ldots(1-q^{-1})}\\
&<&q^{\ell(n-\ell)}\frac{1}{(1-q^{-\ell})(1-q^{-\ell+1})\ldots(1-q^{-1})}\\
&<&q^{\ell(n-\ell)}\prod_{j=1}^\infty\frac{1}{(1-q^{-j})}
\end{eqnarray*}
The function $f(x)=\prod_{j=1}^\infty\frac{1}{(1-x^{j})}$ is
the generating function of integer partitions \cite[Ch.~15]{VLWi01} which
is increasing in $x$.  As we
are interested in $f(1/q)$ for $q\geq 2$, we find that
\[
\prod_{j=1}^\infty\frac{1}{1-q^{-j}} \leq
\prod_{j=1}^\infty\frac{1}{1-2^{-j}}  = 1/Q_0 < 4,
\]
where $Q_0 \approx 0.288788095$ is a probabilistic
combinatorial constant (see, e.g., \cite{Be80}) that gives
the probability that a large, randomly chosen square binary
matrix over $\F_2$ is nonsingular.\qed

We remarked earlier that the Grassmann graph constitutes an
association scheme, which lets us use simple geometric arguments to
give the standard sphere-packing upper bounds and sphere-covering lower
bounds. In order to establish the bounds we need the notion of a
sphere.
  
\begin{defi}\label{def:sphere}
  Let $W$ be an $N$ dimensional vector space and let $\mathcal{P}(W,\ell)$ be the set of $\ell$ dimensional subspaces of $W$.
  The sphere $S(V,\ell,t)$ of radius $t$ centered at a space $V$ in $\mathcal{P}(W,\ell)$ is 
  defined as the set of all subspaces $U$ that satisfy $d(U,V)\leq 2t$,
 \[ S(V,\ell,t)=\{U\in \mathcal{P}(W,\ell):d(U,V)\leq 2t\}.\quad\qed
\]
\end{defi}
Note that we prefer to define the radius in terms of the
graph distance in the Grassmann graph.   The radius can therefore
take on any non-negative integer value.

\begin{thm}\label{thm:spheresize}
The number of spaces in $S(V,\ell,t)$ is independent of $V$ and equals
\[
|S(V,\ell,t)|=\sum_{i=0}^t q^{i^2}\gc{\ell}{i}{}\gc{N-\ell}{i}{}
\]
for $t\leq \ell$.
\end{thm}

\Proof The claim that $S(V,\ell,t)$ is independent of $V$ follows from
the fact that $\mathcal{P}(W,\ell)$ constitutes a distance regular
graph \cite{BCN89}.  We give an expression for the number of spaces
$U$ that intersect $V$ in an $\ell-i$ dimensional subspace. We can
choose the $\ell-i$ dimensional subspace of intersection in
$\gc{\ell}{\ell-i}{} = \gc{\ell}{i}{}$ ways.
Once this is done we can complete the
subspace in
\[
\frac{(q^N-q^\ell)(q^N-q^{\ell+1})\ldots(q^N-q^{\ell+i-1})}{(q^\ell-q^{\ell-i})(q^\ell-q^{\ell-i+1})\ldots(q^\ell-q^{\ell-1})}=q^{i^2}\gc{N-\ell}{i}{}
\]
ways. Thus the cardinality of a shell of spaces at distance $2i$ around $V$
equals $q^{i^2}\gc{N-\ell}{i}{}\gc{\ell}{i}{}$. Summing the cardinality of
the shells gives the theorem.\hfill\qed

Note that $|S(V,\ell,t)| = |S(V,N-\ell,t)|$, as expected
from (\ref{eqn:complement}).

\subsection{Sphere-Packing and Sphere-Covering Bounds}

We now can simply state the sphere-packing and sphere-covering bounds
as follows:

\begin{thm}\label{spherepacking}
Let $\mathcal{C}$ be a collection of spaces in $\mathcal{P}(W,\ell)$ such
that $D(\mathcal{C}) \geq 2t$, and let
$s=\lfloor \frac{t-1}{2} \rfloor$. 
The size of $\mathcal{C}$ must satisfy
\[
|\mathcal{C}|\leq \frac{|\mathcal{P}(W,\ell)|}{|S(V,\ell,s)|}=\frac{\gc{N}{\ell}{}}{|S(V,\ell,s)|}< \frac{\gc{N}{\ell}{}}{q^{s^2}\gc{\ell}{s}{}\gc{N-\ell}{s}{}}<4q^{(\ell-s)(N-s-\ell)}
\]

Conversely, there exists a code $\mathcal{C}'$ with distance $D(\mathcal{C}')\geq 2t$ such that $|\mathcal{C}'|$ is lower bounded by
\begin{eqnarray*}
|\mathcal{C}|\geq \frac{|\mathcal{P}(W,\ell)|}{|S(V,\ell,t-1)|}=\frac{\gc{N}{\ell}{}}{|S(V,\ell,t-1)|}>\frac{\gc{N}{\ell}{}}{(t-1)q^{(t-1)^2}\gc{\ell}t{}\gc{N-\ell}{t-1}{}}>\\
\frac{1}{16t} q^{(\ell-t+1)(N-t-\ell+1)}
\end{eqnarray*}
\end{thm}
\Proof Given the expression for the size of a sphere in $\mathcal{P}(W,\ell)$ the upper and lower bounds are just the familiar packing and covering bounds for codes in distance regular graphs.\hfill\qed

Again we note, as a consequence of (\ref{eqn:complement}),
that these bounds are symmetric in $\ell$ and $N-\ell$.

We can express the bounds of Theorem \ref{spherepacking} in terms of normalized parameters.

\begin{cor}
Let $\mathcal{C}$ be a collection of spaces in $\mathcal{P}(W,\ell)$ with normalized minimum distance $\delta=\frac{D(\mathcal{C})}{2\ell}$.
The rate of $\mathcal{C}$ is bounded from above by
\[
R\leq (1-\delta/2)(1-\lambda(1+\delta/2))+o(1),
\]
where $o(1)$ approaches zero as $N$ grows.
Conversely, there exists a code $\mathcal{C}'$ with normalized distance $\delta$ such that the rate of $\mathcal{C}'$ is lower bounded as:
\[
R\geq (1-\delta)(1-\lambda({\delta}+1))+o(1),
\]
where again $o(1)$ approaches zero as $N$ grows.\hfill\qed
\end{cor}

As in the case of the Hamming scheme, the upper bound is not very good,
especially since it easily can be seen that $\delta$ cannot be larger
than one. We next derive a Singleton type bound for packings in the
Grassmann graph.

\subsection{Singleton Bound}

We begin by defining a suitable puncturing operation on codes.
Suppose $\mathcal{C}$ is a collection of spaces in $\mathcal{P}(W,\ell)$,
where $W$ has dimension $N$. Let
$W'$ be any subspace of $W$ of dimension $N-1$. 
A punctured code $\mathcal{C}'$ is obtained
from $\mathcal{C}$ by replacing each space
$V \in \mathcal{C}$ by $V' = {\cal H}_{\ell-1}(V \cap W')$
where ${\cal H}_{\ell-1}$ denotes the erasure operator defined earlier.
In other words, $V$ is replaced by $V \cap W'$ if $V \cap W'$
has dimension $\ell -1 $;  otherwise $V$ is replaced by some
$(\ell-1)$-dimensional subspace of $V$.   Although this
puncturing operation does not in general result in
a unique code, we denote any such punctured code as $\mathcal{C}|_{W'}$.

We have the following theorem.
\begin{thm}\label{thm:puncturing}
If $\mathcal{C} \subseteq \mathcal{P}(W,\ell)$ is
a code of type $[N,\ell,\log_q|\mathcal{C}|,D]$ with
$D > 2$ and $W'$ is an $(N-1)$-dimensional subspace of $W$,
then
$\mathcal{C}' = \mathcal{C}|_{W'}$
is a code of type
$[N-1,\ell-1,\log_q|\mathcal{C}|,D']$ with $D' \geq D-2$.
\end{thm}

\Proof Only the cardinality and the
minimum distance of $\mathcal{C}'$ are in question. We first verify
that $D' \geq D-2$.  Let $U$ and $V$
be two codewords of $\mathcal{C}$, and suppose that
$U' = {\cal H}_{\ell-1}(U \cap W')$ and
$V' = {\cal H}_{\ell-1}(V \cap W')$ are the corresponding
codewords in $\mathcal{C}'$.
Since $U' \subseteq U$ and $V' \subseteq V$
we have $U' \cap V' \subseteq U \cap V$, so that
$2\dim{U' \cap V'} \leq 2\dim{U\cap V} \leq 2\ell -D$,
where the latter inequality follows
from the property that
$d(U,V) = 2\ell - 2\dim{U \cap V} \geq D$.
Now in $\mathcal{C}'$ 
we have
\begin{eqnarray*}
d(U',V') & = & \dim{U'} + \dim{V'} - 2\dim{U' \cap V'}\\
         & = & 2(\ell-1)-2\dim{U'\cap V'}\\
	 & \geq & 2\ell - 2 - (2\ell - D)\\
	 & = & D-2.
\end{eqnarray*}
Since $D>2$, $d(U',V')>0$, so $U'$ and $V'$ are distinct,
which shows that $C'$ has as many
codewords as $C$.\qed

We may now state the Singleton bound.

\begin{thm}\label{thm:singleton}
  A $q$-ary code of $\mathcal{C} \subseteq \mathcal{P}(W,\ell)$ of
  type $[N,\ell,\log_q|\mathcal{C}|,D]$ must satisfy
  \[
  |\mathcal{C}|
  \leq \gc{N-(D-2)/2}{\max\{\ell,N-\ell\}}{q}. 
  \]
\end{thm}

\Proof  If $\mathcal{C}$ is punctured a total of $(D-2)/2$ times,
a code $\mathcal{C}'$ of type
$[N-(D-2)/2,\ell-(D-2)/2,\log_q|C|,D']$ is obtained, with
every codeword having dimension $\ell-(D-2)/2$ and with $D' \geq 2$.
Such a code cannot have more codewords than the corresponding
Grassmannian, which contains
$a=\gc{N-(D-2)/2}{\ell-(D-2)/2}{q}=\gc{N-(D-2)/2}{N-\ell}{q}$ points.
Applying the same argument to $\mathcal{C}^\perp$ yields
the upper bound $b=\gc{N-(D-2)/2}{\ell}{q}$.
Now $a < b$ if and only if $\ell < N - \ell$, from which
the bound follows.  \qed

This bound is easily expressed in terms of normalized parameters.
We consider only the case where $\ell \leq N-\ell$, i.e.,
$\lambda \leq 1/2$.

\begin{cor}\label{cor:singleton}
Let $\mathcal{C}$ be a collection of spaces in $\mathcal{P}(W,\ell)$,
with $\ell \leq \dim{W}/2$ and
with normalized minimum distance $\delta=\frac{D(\mathcal{C})}{2\ell}$.
The rate of $\mathcal{C}$ is bounded from above by
\[
R\leq (1-\delta)(1-\lambda)+\frac{1}{\lambda N}(1-\lambda+o(1)).
\]
\end{cor}

\begin{figure}[htbp]
\centerline{\input{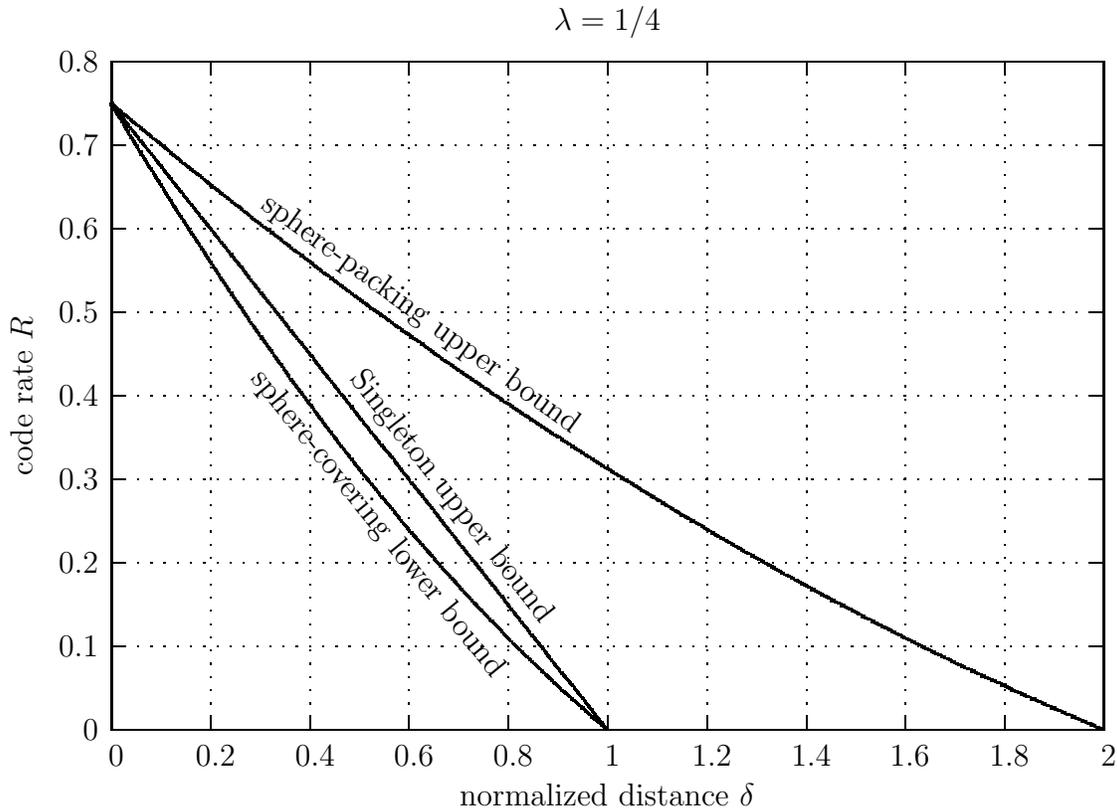}}
\caption{Upper and lower asymptotic bounds on the largest rate of a code in 
the Grassmann graph $G_{W,\ell}$ where the
dimension $N$ of ambient vector space 
is asymptotically large and $\lambda=\frac{\ell}{N}$ is chosen as 1/4.}
\label{fig:bounds}
\end{figure}

The three bounds are depicted in Fig.~\ref{fig:bounds},
for $\lambda = 1/4$ and in the limit as $N \rightarrow \infty$.

\section{A Reed-Solomon-like Code Construction and Decoding Algorithm}
\label{sec:RS}

We now turn to the problem of constructing
a code capable of correcting errors and erasures
at the output of the operator channels defined in Section~\ref{sec:setup}.
The code construction is equivalent to that 
given by Wang, Xing and Safavi-Naini \cite{WXSN03} in
the context of linear authentication codes,
which in turn can be regarded as an application of the
maximum rank-distance construction of Gabidulin \cite{Ga85}.
(The connection between constant-dimension codes
and rank-metric codes and the generalized rank-metric decoding
problem induced by the operator channel is studied
in detail in \cite{SiKsKo07}.)
The main contribution of this section is
a Sudan-style ``list-1'' minimum distance decoding algorithm,
given in Section~\ref{subsec:decoding}.

\subsection{Linearized Polynomials}
\label{subsec:linearized}

Let $\F_q$ be a finite field and let $\F = \F_{q^m}$ be an extension field.
Recall from \cite[Ch.~11]{Be68}, \cite[Sec.~4.9]{MaSl77},
\cite[Sec.~3.4]{LiNi83} that a polynomial $L(x)$
is called a \emph{linearized polynomial}
over $\F$ if it takes the form
\begin{equation}
L(x) = \sum_{i=0}^d a_i x^{q^i},
\label{eqn:linearized}
\end{equation}
with coefficients $a_i \in \F, i=0, \ldots, d$.
If all coefficients are zero, so that $L(x)$ is the zero
polynomial, we will write $L(x) \equiv 0$;  more generally,
we will write $L_1(x) \equiv L_2(x)$ if $L_1(x)-L_2(x) \equiv 0$.
When $q$ is fixed under discussion, we will let $x^{[i]}$
denote $x^{q^i}$.  In this notation, a linearized polynomial
over $\F$ may be written as
\[
L(x) = \sum_{i=0}^d a_i x^{[i]}.
\]

If $L_1(x)$ and $L_2(x)$ are linearized polynomials over $\F$,
then so is any $\F$-linear combination $\alpha_1 L_1(x) + \alpha_2 L_2(x)$,
$\alpha_1, \alpha_2 \in F$.  The ordinary
product $L_1(x) L_2(x)$ is not necessarily a linearized polynomial.
However, the composition $L_1(L_2(x))$, often
written as $L_1(x) \otimes L_2(x)$, of two linearized
polynomials over $\F$ is again a linearized polynomial
over $\F$.  Note that this operation is not commutative,
i.e., $L_1(x) \otimes L_2(x)$ need not be equal to $L_2(x) \otimes L_1(x)$.

The product $L_1(x) \otimes L_2(x)$ of linearized polynomials is computed
explicitly as follows.
If $L_1(x) = \sum_{i \geq 0} a_i x^{[i]}$
and $L_2(x) = \sum_{j \geq 0} b_j x^{[j]}$, then
\begin{eqnarray*}
L_1(x) \otimes L_2(x) & = & L_1(L_2(x) = \sum_{i \geq 0} a_i (L_2(x))^{[i]}\\
& = & \sum_{i \geq 0} a_i \left( \sum_{j \geq 0} b_j x^{[j]} \right)^{[i]}\\
& = & \sum_{i \geq 0} \sum_{j \geq 0} a_i b_j^{[i]} x^{[i+j]}
 =  \sum_{k \geq 0} c_k x^{[k]}
\end{eqnarray*}
where
\[
c_k = \sum_{i=0}^k a_i b_{k-i}^{[i]}.
\]
Thus the coefficients of $L_1(x) \otimes L_2(x)$ are obtained
from those of $L_1(x)$ and $L_2(x)$ via a modified convolution operation.
If $L_1(x)$ has degree $q^{d_1}$ and $L_2(x)$ has 
degree $q^{d_2}$, then both $L_1(x) \otimes L_2(x)$ and $L_2(x) \otimes L_1(x)$
have degree $q^{d_1 + d_2}$.

Under addition + and composition $\otimes$, the set of linearized
polynomials over $\F$ forms a non-commutative ring with identity.
Although non-commutative, this ring has many of the properties of
a Euclidean domain including,  for example, an absence of zero-divisors.
The degree of a nonzero element forms a natural norm.
There are two division algorithms:
a left division and a right division, i.e., given any two linearized
polynomials $a(x)$ and $b(x)$, it is easy
to prove by induction that there exist unique linearized polynomials
$q_L(x)$, $q_R(x)$, $r_L(x)$ and $r_R(x)$ such that
\[
a(x) = q_L(x) \otimes b(x) + r_L(x) = b(x) \otimes q_R(x) + r_R(x),
\]
where $r_L(x) \equiv 0$ or $\deg(r_L(x)) < \deg(b(x))$ and similarly
where $r_R(x) \equiv 0$ or $\deg(r_R(x)) < \deg(b(x))$.

The polynomials $q_R(x)$ and $r_R(x)$
are easily determined by the following straightforward
variation of ordinary polynomial long division.
Let $\lc(a(x))$ denote the leading coefficient of $a(x)$,
so that if $a(x)$ has degree $q^d$, i.e.,
$a(x) = a_d x^{[d]} + a_{d-1} x^{[d-1]} + \cdots + a_0 x^{[0]}$
with $a_d \neq 0$, then ${\lc}(a(x)) = a_d$.

\begin{tabbing}
{\tt\ \ \ }\={\tt\ \ \ }\={\tt\ \ \ }\={\tt\ \ \ }\kill
\>{\tt procedure} {\tt RDiv}($a(x),b(x)$)\\
\>\>{\tt input}: a pair $a(x),b(x)$ of linearized polynomials over $\F = \F_q^m$, with $b(x) \not\equiv 0$.\\
\>\>{\tt output}: a pair $q(x),r(x)$ of linearized polynomials over $\F_q^m$\\
\>{\tt begin}\\
\>\> {\tt if} $\deg(a(x)) < \deg(b(x))$ {\tt then}\\
\>\> \> {\tt return} $(0, a(x))$\\
\>\> {\tt else}\\
\>\> \> $d := \deg(a(x))$, $e := \deg(b(x))$, $a_d := \lc(a(x))$, $b_e := \lc(b(x))$ \= \\
\>\> \> $t(x) := (a_d/b_e)^{[m-e]} x^{[d-e]}$ \> (*)\\
\>\> \> {\tt return} $(t(x),0) + {\tt RDiv}(a(x)-b(x)\otimes t(x),b(x))$ \>(**)\\
\>\> {\tt endif}\\
\>{\tt end}
\end{tabbing}

Note that the parameter $m$ in step (*) is equal to the dimension
of $\F_q^m$ as a vector space over $\F_q$.
This algorithm terminates when it produces polynomials $q(x)$ and
$r(x)$ with the property that $a(x) = b(x) \otimes q(x) + r(x)$
and either $r(x) \equiv 0$ or $\deg r(x) < \deg b(x)$.

The left-division procedure is essentially the same;
``{\tt RDiv}'' is replaced with ``{\tt LDiv}''
and (*) and (**) are replaced with the following:
\begin{tabbing}
{\tt\ \ \ }\={\tt\ \ \ }\={\tt\ \ \ }\={\tt\ \ \ }\kill
\>\>\> $t(x) := (a_d/(b_e^{[d-e]})) x^{[d-e]}$ \\
\>\>\> {\tt return} $(t(x),0) + {\tt LDiv}(a(x)-t(x) \otimes b(x),b(x))$
\end{tabbing}
With this change, the algorithm terminates when it produces
polynomials $q(x)$ and $r(x)$ with the property that
$a(x) = q(x) \otimes b(x) + r(x)$.

Linearized polynomials receive their name from the
following property.  Let $L(x)$ be a linearized polynomial
over $\F$, and
let $K$ be an arbitrary extension field of $\F$.
Then $K$ may be regarded as a vector space over $\F_q$.
The map taking $\beta \in K$ to $L(\beta) \in K$ is
\emph{linear} with respect to $\F_q$, i.e.,
for all $\beta_1,\beta_2 \in K$ and all $\lambda_1,\lambda_2 \in \F_q$,
\[
L(\lambda_1 \beta_1 + \lambda_2 \beta_2) = \lambda_1 L(\beta_1)
+ \lambda_2 L(\beta_2).
\]

Suppose that $K$ is chosen to be large enough to include all the zeros
of $L(x)$.
The zeros of $L(x)$ then correspond to the kernel of
$L(x)$ regarded as a linear map, so they
form a vector space over $\F_q$.  If $L(x)$ has degree $q^d$, this
vector space has dimension at most $d$,
but the dimension could possibly be smaller if $L(x)$ has repeated
roots (which occurs if and only if $a_0 = 0$ in (\ref{eqn:linearized})).

On the other hand
if $V$ is an $n$-dimensional subspace of $K$, then
\[
L(x) = \prod_{\beta \in V} (x - \beta)
\]
is a monic linearized polynomial over $K$ (though not necessarily
over $\F$).
See \cite[Lemma 21]{MaSl77} or \cite[Theorem 3.52]{LiNi83}.

The following lemma shows that if two linearized polynomials
of degree at most $q^{d-1}$ agree on at least $d$ linearly independent
points, then the two polynomials coincide.
\begin{lem}\label{lem:coincident}
Let $d$ be a positive integer and
let $f(x)$ and $g(x)$ be two linearized polynomials over $\F$
of degree less than $q^d$. If
$\alpha_1, \alpha_2, \ldots, \alpha_d$
are linearly independent elements of $K$ such that
have $f(\alpha_i) = g(\alpha_i)$ for $i = 1,\ldots,d$,
then $f(x) \equiv g(x)$.
\end{lem}

\Proof
Observe that $h(x) = f(x)-g(x)$
has $\alpha_1, \ldots, \alpha_d$ as zeros, and hence also has all $q^d$
linear combinations of these elements as zeros.  Thus $h(x)$
has at least $q^{d}$ distinct zeros.  However, since the actual degree
of $h(x)$ is strictly smaller than $q^d$, this is only possible
if $h(x) \equiv 0$.\qed

\subsection{Code Construction}

Just as traditional Reed-Solomon codeword components
may be obtained via the evaluation of an \emph{ordinary} message polynomial,
we obtain here a basis for the transmitted vector
space via the evaluation of a \emph{linearized} message polynomial.

Let $\F_q$ be a finite
field, and let $\F = \F_{q^m}$ be a (finite) extension field of $\F_q$.
As in the previous subsection, we may regard $\F$
as a vector space of dimension $m$ over $\F_q$.
Let $A = \{ \alpha_1, \ldots, \alpha_\ell \} \subset \F$ be a set
of linearly independent elements in this vector space.
These elements span an $\ell$-dimensional vector space $\langle A \rangle \subseteq \F$ over $\F_q$.  Clearly $\ell \leq m$.
We will take as ambient space the direct sum
$W = \langle A \rangle \oplus \F =
\{ (\alpha,\beta): \alpha\in \langle A \rangle, \beta \in \F \}$,
a vector space of dimension $\ell + m$ over $\F_q$.

Let $u=(u_0, u_1, \ldots, u_{k-1}) \in \F^k$ denote a block
of message symbols, consisting of $k$ symbols over $\F$ or,
equivalently, $mk$ symbols over $\F_q$.  Let $\F^k[x]$
denote the set of linearized polynomials over $\F$ of degree
at most $q^{k-1}$.
Let $f(x) \in \F^k[x]$, defined as
\[
f(x) = \sum_{i=0}^{k-1} u_i x^{[i]},
\]
be the linearized polynomial with coefficients corresponding to $u$.
Finally, let $\beta_i = f(\alpha_i)$.
Each pair $(\alpha_i,\beta_i)$, $i = 1, \ldots, \ell$,
may be regarded as a vector in $W$.
Since $\{\alpha_1,\ldots,\alpha_\ell\}$ is a linearly
independent set, so is
$\{ (\alpha_1,\beta_1),\ldots,(\alpha_\ell,\beta_\ell)\}$;
hence this set spans an $\ell$-dimensional subspace $V$ of $W$.
We denote the map that takes the message polynomial $f(x) \in \F^k[x]$ to
the linear space $V \in \mathcal{P}(W,|A|)$ as ${\ev}_A$.

\begin{lem}\label{lem:injection}
If $|A| \geq k$ then the map
${\ev}_A: \F^k[x] \rightarrow \mathcal{P}(W,|A|)$
is injective.
\end{lem}

\Proof  Suppose $|A|\geq k$ and ${\ev}_A(f) = {\ev}_A(g)$
for some $f(x),g(x) \in \F^k[x]$.  Let $h(x) = f(x)-g(x)$.
Clearly $h(\alpha_i)=0$ for $i=1,\ldots,\ell$.  Since $h(x)$
is a linearized polynomial,
it follows that $h(x) = 0$
for all $x \in \langle A \rangle$.  Thus $h(x)$ has at
least $q^{|A|} \geq q^k$ zeros,
which is only possible (since $h(x)$ has 
degree at most $q^{k-1}$)  if $h(x) \equiv 0$, so that
$f(x) \equiv g(x)$.\qed

Henceforth we will assume that $\ell \geq k$.
Lemma~\ref{lem:injection} implies that, provided this condition is satisfied,
the image of $\F^k[x]$ is a code $\mathcal{C} \subseteq \mathcal{P}(W,\ell)$
with $q^{mk}$ codewords.
The minimum distance of $\mathcal{C}$ is
given by the following theorem; however, first we need the following lemma.

\begin{lem}\label{lem:indep}
If $\{ (\alpha_1,\beta_1), \ldots, (\alpha_r,\beta_r)\} \subseteq W$
is a collection of $r$ linearly independent elements
satisfying $\beta_i = f(\alpha_i)$ for
some linearized polynomial $f$ over $\F$,
then $\{ \alpha_1,\ldots,\alpha_r \}$ is a linearly independent set.
\end{lem}
\Proof Suppose that
for some $\gamma_1,\ldots,\gamma_r \in \F_q$ we have
$\sum_{i=1}^r \gamma_i \alpha_i = 0$.
Then, in $W$, we would have
\begin{eqnarray*}
\sum_{i=1}^r \gamma_i (\alpha_i,\beta_i) &  =  &
\left(\sum_{i=1}^r\gamma_i \alpha_i,\sum_{i=1}^r\gamma_i \beta_i\right) 
 =  \left(0,\sum_{i=1}^r\gamma_i f(\alpha_i)\right) 
 =  \left(0,f\left(\sum_{i=1}^r \gamma_i \alpha_i\right)\right) \\
 & =  & (0,f(0)) = (0,0),
\end{eqnarray*}
which is possible (since the $(\alpha_i,\beta_i)$ pairs
are linearly independent)
only if $\gamma_1,\ldots,\gamma_r = 0$.\qed

\begin{thm}\label{thm:RSparms}
Let $\mathcal{C}$ be the image under ${\ev}_A$ of $\F^k[x]$,
with $\ell = |A| \geq k$.
Then $\mathcal{C}$ is a code of type $[ \ell + m, \ell, mk, 2(\ell-k+1)]$.
\end{thm}

\Proof Only the minimum distance is in question.  Let $f(x)$
and $g(x)$ be distinct elements of $\F^k[x]$, and let
$U = {\ev}_A(f)$ and $V={\ev}_A(g)$.  Suppose that
$U \cap V$ has dimension $r$.  This means it is possible to
find $r$ linearly independent elements
$(\alpha_1',\beta_1'),\ldots,(\alpha_r',\beta_r')$ such
that $f(\alpha_i') = g(\alpha_i') = \beta_i'$, $i=1,\ldots,r$.
By Lemma~\ref{lem:indep},
$\alpha_1', \ldots, \alpha_r'$ are linearly independent
and hence they span
an $r$-dimensional space $B$ with the property
that $f(b) - g(b) = 0$ for all $b \in B$.
If $r \geq k$ then $f(x)$ and $g(x)$ would be two 
linearized polynomials of degree less than $q^k$
that agree on at least $k$ linearly independent points,
and hence by Lemma~\ref{lem:coincident}, we would have
$f(x) \equiv g(x)$.  Since this is not the case,
we must have $ r \leq k-1$.  Thus
\[
d(U,V) = \dim{U}+\dim{V}-2\dim{U\cap V} = 2(\ell - r) \geq 2(\ell-k+1).
\]
It is easy to exhibit two codewords $U$ and $V$ that
satisfy this bound with equality.
\qed

The Singleton bound, evaluated for the code parameters
of Theorem~\ref{thm:RSparms},
states that
\[
|\mathcal{C}| \leq \gc{N-(D-2)/2}{\ell-(D-2)/2}{q} = \gc{m+k}{k}{q}
< 4 q^{mk}.
\]
This implies that
a true Singleton-bound-achieving code could have
no more than 4 times as many codewords as $\mathcal{C}$.
When $N$ is large enough, the difference in rate
between $\mathcal{C}$ and a Singleton-bound-achieving becomes
negligible.  Indeed,
in terms of normalized parameters, we have
\[
R = (1-\lambda)(1-\delta + \frac{1}{\lambda N})
\]
which certainly has the same asymptotic behavior as the
Singleton bound in the limit as $N \rightarrow \infty$.
We claim, therefore, that these Reed-Solomon-like
codes are nearly Singleton-bound-achieving.

We note also that the traditional network code $\mathcal{C}$
of Example~\ref{exam:header}, a code
of type $[m+\ell,\ell,m\ell,2]$, is obtained
as a special case of these codes by setting $k=\ell$.

This code construction involving the evaluation of
linearized polynomials is clearly closely related to the
rank-metric code construction of Gabidulin \cite{Ga85}.
However, in our setup, the codewords are not
arrays, but rather the vector spaces spanned by the
rows of the array, and the relevant decoding metric
is not the rank metric, but rather the distance
measure defined in (\ref{eqn:distance}).
The connection between subspace codes
and rank-metric codes is explored further
in \cite{SiKsKo07}.

\subsection{Decoding}
\label{subsec:decoding}

Suppose now that $V \in \mathcal{C}$ is transmitted
over the operator channel described in Section~\ref{sec:setup} and
that an $(\ell - \rho + t)$-dimensional subspace $U$ of $W$ is received,
where $\dim{U \cap V} = \ell - \rho$.   In this situation, we have
$\rho$ erasures and an error norm of $t$, and $d(U,V) = \rho + t$.
We expect to be able to recover $V$ from $U$ provided
that $\rho + t < D/2 = \ell -k + 1$, and we will describe
a Sudan-style ``list-1'' minimum distance decoding algorithm to do so
(see, e.g., \cite[Sec.~9.3]{roth}).
Note that, even if $t=0$, we require
$\rho < \ell - k + 1$, or $\ell - \rho \geq k$,
i.e., not surprisingly (given
that we are attempting to recover
$mk$ information symbols),
the receiver must collect enough vectors to span a space of
dimension at least $k$.


Let $r = \ell - \rho + t$ denote the dimension of the
received space $U$, and
let $(x_i,y_i)$, $i=1,\ldots, r$ be a basis for $U$.
At the decoder we suppose that
it is possible to construct a nonzero bivariate polynomial $Q(x,y)$ of the
form
\begin{equation}
Q(x,y) = Q_x(x) + Q_y(y), \mbox{ such that }
Q(x_i,y_i) = 0 \mbox { for } i=1,\ldots,r,
\label{eqn:defQ}
\end{equation}
where $Q_x(x)$ is a linearized polynomial over $\F_{q^m}$
of degree at most $q^{\tau-1}$
and $Q_y(y)$ is a linearized
polynomial over $\F_{q^m}$ of degree at most $q^{\tau-k}$.
Although $Q(x,y)$ is chosen to interpolate only a basis
for $U$,
since $Q(x,y)$ is a linearized polynomial, it follows that
in fact $Q(x,y)=0$ for all $(x,y) \in U$.

We note that (\ref{eqn:defQ}) defines a homogeneous system
of $r$ equations in $2\tau-k+1$ unknown coefficients.  This system
has a nonzero solution when it is under-determined, i.e., when
\begin{equation}
r = \ell- \rho + t < 2\tau - k + 1.
\label{eqn:deccon1}
\end{equation}

Since $f(x)$ is a linearized polynomial over $\F_{q^m}$,
so is $Q(x,f(x))$, given by
\[
Q(x,f(x))  =  Q_x(x) + Q_y(f(x))  =  Q_x(x) + Q_y(x) \otimes f(x).
\]
Since the degree of $f(x)$ is
at most $q^{k-1}$, the degree of $Q(x,f(x))$ is at most $q^{\tau-1}$.

Now let $\{ (a_1,b_1), \ldots, (a_{\ell-\rho},b_{\ell-\rho})\}$
be a basis for $U \cap V$.  Since all vectors of $U$ are zeros of $Q(x,y)$,
we have $Q(a_i,b_i)=0$ for $i=1,\ldots,\ell-\rho$.
However, since $(a_i,b_i) \in V$
we also have $b_i=f(a_i)$ for $i=1,\ldots,\ell-\rho$.
In particular,
\[
Q(a_i,b_i) = Q(a_i,f(a_i)) = 0,~ i = 1,\ldots,\ell-\rho,
\]
thus $Q(x,f(x))$ is a linearized polynomial having
$a_1, \ldots, a_{\ell-\rho}$ as roots.
By Lemma~\ref{lem:indep},
these roots are linearly independent.
Thus $Q(x,f(x))$ is a linearized polynomial of 
degree at most $q^{\tau-1}$ that evaluates to
zero on a space of dimension $\ell-\rho$.
If the condition
\begin{equation}
\ell-\rho \geq \tau
\label{eqn:deccon2}
\end{equation}
holds, then $Q(x,f(x))$ has more zeros than its degree,
which is only possible if
$Q(x,f(x)) \equiv 0$.
Since in general
\[
Q(x,y) = Q_y(y-f(x)) + Q(x,f(x)),
\]
we have, when $Q(x,f(x))\equiv 0$,
\[
Q(x,y) = Q_y(y-f(x))
\]
and so we may hope to extract $y-f(x)$ from $Q(x,y)$.
Equivalently, we may hope to find $f(x)$
from the equation
\begin{equation}
Q_y(x) \otimes f(x) + Q_x(x) \equiv 0.
\label{eqn:key}
\end{equation}
However, this equation is easily solved
using the {\tt RDiv} procedure described in
Section~\ref{subsec:linearized}, with $a(x) = -Q_x(x)$ and
$b(x) = Q_y(x)$.  Alternatively, we can expand 
(\ref{eqn:key}) into a system of equations involving
the unknown coefficients of $f(x)$; this system is readily solved
recursively (i.e., via back-substitution).

In summary, to find nonzero $Q(x,y)$ we must
satisfy (\ref{eqn:deccon1}) and to ensure
that $Q(x,f(x))\equiv 0$ we must satisfy
(\ref{eqn:deccon2}).  When both (\ref{eqn:deccon1}) and (\ref{eqn:deccon2})
hold for some $\tau$, we say that the received space $U$
is \emph{decodable}.

Suppose that the received space $U$ is decodable.
Substituting (\ref{eqn:deccon2}) into (\ref{eqn:deccon1}),
we obtain the condition
$\ell- \rho + t < 2(\ell - \rho) - k + 1$ or, equivalently,
\begin{equation}
\rho + t < \ell - k + 1,
\label{eqn:deccon3}
\end{equation}
i.e., not surprisingly decodability implies (\ref{eqn:deccon3}).

Conversely, suppose (\ref{eqn:deccon3}) is satisfied.
From (\ref{eqn:deccon3}) we get
$\ell - \rho \geq t + k$, or
\begin{equation}
\ell - \rho + t + k  = r+k \leq  2(\ell - \rho).
\label{eqn:equivdeccon3}
\end{equation}
By selecting 
\[
\tau = \lceil \frac{r+k}{2} \rceil
\]
(which is possible to do since the receiver knows both $r$ and $k$),
we satisfy (\ref{eqn:deccon1}).
With this choice of $\tau$, and applying condition (\ref{eqn:equivdeccon3}),
we see that
\[
\tau \leq \ell -\rho + 1/2;
\]
however, since $\ell$, $\rho$ and $\tau$ are integers,
we see that (\ref{eqn:deccon2}) is also satisfied.
In other words, condition (\ref{eqn:deccon3}) implies decodability,
which is precisely what we would have hoped for.

The interpolation polynomial $Q(x,y)$ can be obtained
from the $r$ basis vectors
$(x_1,y_1)$, $(x_2,y_2)$, \ldots, $(x_r,y_r)$ for $U$
via any method that provides a nonzero solution
to the homogeneous system (\ref{eqn:defQ}). We next describe
an efficient algorithm to accomplish this task.
This algorithm is closely related to the work
of Loidreau \cite{Lo06}, who provides a polynomial-reconstruction-based
procedure for rank-metric decoding of Gabidulin codes.

Let $f(x,y)=f_x(x)+f_y(y)$ be a bivariate linearized polynomial, which means 
that both $f_x(x)$ and $f_y(y)$ are linearized polynomials.   
Let the degree of $f_x(x)$ and $f_y(y)$  be
$q^{d_x(f)}$ and $q^{d_y(f)}$, respectively. 
The $(1,k-1)$-weighted degree of $f(x,y)$ is defined as 
\[ \deg_{1,k-1}(f(x,y)):= \max\{d_x(f), k-1+ d_y(f)\}\]
Note that this definition is different from the weighted degree definitions
for usual bivariate polynomials. However, it should become  more natural by
observing that we may write $f(x,y)$ as $f(x,y)=f_x(x)+f_y(x)\otimes y$.

The following adaptation of an algorithm for the 
interpolation problem in Sudan-type decoding algorithms 
(see e.g. \cite{McEliece1,McEliece2}) 
provides an efficient way to find the required bivariate
linearized polynomial
$Q(x,y)$. Let a vector space $U$ be spanned by $r$
linearly independent points $(x_i,y_i)\in W$.

\begin{tabbing}
{\tt\ \ \ }\={\tt\ \ \ }\={\tt\ \ \ }\={\tt\ \ \ }\={\tt\ \ \ }\={\tt\ \ \ }\=\kill
\>{\tt procedure} {\tt Interpolate}$(U)$\\
\>\>{\tt input}: a basis $(x_i,y_i)\in W$, $i = 1, \ldots, r$, for $U$\\
\>\>{\tt output:} a linearized bivariate polynomial $Q(x,y) = Q_x(x)+Q_y(y)$\\
\>\>{\tt initialization}: $f_0(x,y) = x$, $f_1(x,y) = y$\\
\>{\tt begin}\\
\>\> {\tt for} $i=1$ {\tt to} $r$ {\tt do}\\
\>\> \> $\Delta_0 := f_0(x_i,y_i)$; $\Delta_1 := f_1(x_i,y_i)$\\
\>\> \> {\tt if} $\Delta_0 = 0$ {\tt then}\\
\>\> \> \> $f_1(x,y) := f_1^q(x,y)-\Delta_1^{q-1}f_1(x,y)$\\
\>\> \> {\tt elseif} $\Delta_1 = 0$ {\tt then}\\
\>\> \> \> $f_0(x,y) := f_0^q(x,y)-\Delta_0^{q-1}f_0(x,y)$\\
\>\> \> {\tt else}\\
\>\> \> \> {\tt if} $\deg_{1,k-1}(f_0) \leq \deg_{1,k-1}(f_1)$ {\tt then}\\
\>\>\>\>\> $f_1(x,y) := \Delta_1 f_0(x,y) - \Delta_0 f_1(x,y)$\\
\>\>\>\>\> $f_0(x,y) := f_0^q(x,y)-\Delta_0^{q-1}f_0(x,y)$\\
\>\> \> \> {\tt else}\\
\>\>\>\>\> $f_0(x,y) := \Delta_1 f_0(x,y) - \Delta_0 f_1(x,y)$\\
\>\>\>\>\> $f_1(x,y) := f_1^q(x,y)-\Delta_1^{q-1}f_1(x,y)$\\
\>\> \> {\tt endif}\\
\>\> {\tt endfor}\\
\>\> {\tt if} $\deg_{1,k-1}(f_1)<\deg_{1,k-1}(f_0)$ {\tt then}\\
\>\> \> {\tt return} $f_1(x,y)$\\
\>\> {\tt else}\\
\>\> \> {\tt return} $f_0(x,y)$\\
\>\> {\tt endif}\\
\>{\tt end}
\end{tabbing}

For completeness we provide a proof of correctness of this algorithm, which
mimics the proof in the case of standard bivariate interpolation,
finding the ideal of polynomials that vanishes at a given set of
points \cite{McEliece1,McEliece2}.

Define an order $\prec$ on bivariate linearized polynomials as
follows:
write $f(x,y) \prec g(x,y)$ if
$\deg_{1,k-1}(f(x,y))<\deg_{1,k-1}(g(x,y))$. 
In case $\deg_{1,k-1}(f(x,y))=\deg_{1,k-1}(g(x,y))$ write
$f(x,y)\prec g(x,y)$ if $d_y(f)+k-1<\deg_{1,k-1}(f(x,y))$ and
$d_y(g)+k-1=\deg_{1,k-1}(g(x,y))$.
If none of these conditions is true, we say
that $f(x,y)$ and $g(x,y)$ are not comparable. 
While $\prec$ is
clearly not a total order on polynomials,
it is granular enough for the proof of correctness of procedure
{\tt Interpolate}.
In particular, $\prec$ gives a total order on
monomials and we can, hence, define  a leading term $\lt_\prec(f)$ as the
maximal monomial (without its coefficient) in $f$  under the order $\prec$.

\begin{lem}
Assume that we have two bivariate linearized polynomials  $f(x,y)$  and
$g(x,y)$ which are not comparable under $\prec$. 
We  can create a linear combination $h(x,y)=f(x,y)+\gamma g(x,y)$ which for a
suitably chosen $\gamma$ yields a polynomial
$h(x,y)$ with $h(x,y)\prec f(x,y)$ and $h(x,y)\prec g(x,y)$.
\end{lem}

\Proof If $f(x,y)$ and $g(x,y)$ are not comparable then we have
$\lt_\prec(f)=\lt_\prec(g)$. Choosing $\gamma$ as the quotient of the
corresponding coefficients of $\lt_\prec(f)$ in $f$ and $\lt_\prec(g)$ in $g$
yields a polynomial $h(x,y)$ such that $\lt_\prec(h)\prec
\lt_\prec(f)=\lt_\prec(g)$.\qed

Let $A$ be a set of $r$ linearly independent points $(x_i,y_i)\in W$.
We say that a nonzero polynomial $f(x,y)$ is $x$-minimal with respect
to $A$ if $f(x,y)$ is a minimal polynomial under $\prec$ such
that $\lt_\prec(f)=x^{[d_x(f)]}$ and 
$f(x,y)$ vanishes at all points of $A$. Similarly, a nonzero
polynomial $g(x,y)$
is said to be $y$-minimal with respect to $A$ if $g(x,y)$ is a a minimal 
polynomial under $\prec$ such that $\lt_\prec(g)=y^{[d_y(g)]}$ and 
$g(x,y)$ vanishes in all points of $A$.

\begin{thm}\label{Interpolation}
  The polynomials $f_0(x,y)$ and $f_1(x,y)$ that are output by procedure
  {\tt Interpolate}
  are $x$-minimal and $y$-minimal with respect to the given
  set of $r$ linearly independent points $(x_i,y_i)\in W$.  
\end{thm}

\Proof First we note  that
  $x$-minimal and $y$-minimal polynomials can always be compared under
  $\prec$ since they have different leading monomials.
  The proof proceeds by induction. We first verify that the polynomials
  $x$ and $y$ are $x$-minimal and $y$-minimal with respect to the empty set.
  We thus assume that after $j$ iterations of the interpolation algorithm the
  polynomials $f_0(x,y)$ and $f_1(x,y)$ are $x$-minimal and $y$-minimal with
  respect to the points $(x_i,y_i)$, $i=1,2,\ldots,j$. It is easy to check
  that the set of polynomials constructed in the next iteration also vanishes
  at the point $(x_{j+1},y_{j+1})$ so this part of the definition of $x$- and
  $y$-minimality with respect to points $(x_i,y_i)$, $i=1,2,\ldots,j+1$ will
  not be a problem. 

  Assume first the generic case that $\Delta_0\neq 0$ and $\Delta_1\neq 0$.
  Assume that $f_1(x,y)\prec f_0(x,y)$ holds. Let $f_0'(x,y)=\Delta_1 f_0(x,y)
  - \Delta_0 f_1(x,y)$.  In this case $lt_\prec(f_0'(x,y))=lt_\prec(f_0(x,y))$
  and the $x$-minimality of $f_0'(x,y)$ follows from the $x$-minimality of
  $f_0(x,y)$. Let $f'_1(x,y)= f_1^q(x,y)-\Delta_1^{q-1}f_1(x,y)$. We will show
  that $f'_1(x,y)$ is $y$-minimal with respect to points $(x_i,y_i)$,
  $i=1,2,\ldots,j+1$. To this end and in order to arrive at a contradiction assume that
  $f'_1(x,y)$ is not $y$-minimal. This would imply that there exists a $y$-minimal
  polynomial $f_1''(x,y)$ w.r.t. points $(x_i,y_i)$,
  $i=1,2,\ldots,j+1$ such that $f_1''(x,y)\neq f_1(x,y)$ which has the same leading term as
  $f_1(x,y)$. The two polynomials are clearly different since $f_1''(x,y)$
  would vanish at $(x_{j+1},y_{j+1})$ while $f_1(x,y)$ does not. But this
  would imply that we can find a polynomial $h(x,y)$ as linear combination of
  $f_1''(x,y)$ and $f_1(x,y)$ which would precede both $f_0(x,y)$ and
  $f_1(x,y)$ under the order $\prec$ and which would vanish at all points $(x_i,y_i)$,
  $i=1,2,\ldots,j$, thus contradicting the minimality of $f_0(x,y)$ and
  $f_1(x,y)$. A virtually identical arguments holds if we have
  $f_0(x,y)\prec f_1(x,y)$.  

  Next we consider the case that $\Delta_0$ equals $0$ while we have
  $\Delta_1\neq 0$.  In this case $f_0(x,y)$ is unchanged and, hence, inherits
  its $x$-minimality from the previous iteration.  We only have to check that
  the newly constructed $f'_1(x,y)= f_1^q(x,y)-\Delta_1^{q-1}f_1(x,y)$ is
  $y$-minimal with respect to points $(x_i,y_i)$, $i=1,2,\ldots,j+1$. Again,
  assuming the opposite would imply that there exists a polynomial
  $f_1''(x,y)\neq f_1(x,y)$ with the same leading term as $f_1(x,y)$. The two
  polynomials are again different since $f_1''(x,y)$ would vanish at
  $(x_{j+1},y_{j+1})$ while $f_1(x,y)$ does not. Again, we form a suitable
  linear combination $h(x,y)$ of $f_1''(x,y)$ and $f_1(x,y)$ which would
  precede $f_1(x,y)$ under $\prec$. If the leading term of $h(x,y)$ is of type
  $y^{[d_y(h)]}$ or $h(x,y)\prec f_0(x,y)$ we have arrived at a contradiction
  negating the $y$-minimality of $f_1(x,y)$ or the $x$-minimality of
  $f_0(x,y)$. Otherwise, note that $h(x,y)$ does not vanish at
  $(x_{j+1},y_{j+1})$ and hence is not a multiple (under $\otimes$ of
  $f_0(x,y)$.)  Hence, for a suitably chosen $t$, we can find a polynomial
  $h''(x,y)$ as a linear combination of $h(x,y)$ and $x^{[t]}\otimes f_0(x,y)$
  which precedes $h(x,y)$.  Repeating this procedure we arrive at a polynomial
  $\hat{h}(x,y)$ which either has a leading term of type $y^{[d_y(\hat{h})]}$
  or which precedes $f_0(x,y)$ under $\prec$, either contradicting the
  $y$-minimality of $f_1(x,y)$ or the $x$-minimality of $f_0(x,y)$.  Finally
  we note that the case $\Delta_0\neq 0$ and $\Delta_1=0$ follows from
  similar arguments. For the case $\Delta_0=0$ and $\Delta_1=0$ there is
  nothing to prove.\qed

Based on Theorem \ref{Interpolation} we can claim that the {\tt Interpolate}
procedure solves the problem---as required
in equation (\ref{eqn:defQ})---of finding the bivariate linearized
polynomial $Q(x,y)$ of minimal $(1,k-1)$ weighted degree $\tau-1$,
which is identified as the polynomial $f_0(x,y)$ or
$f_1(x,y)$ of smaller $(1,k-1)$ weighted degree. 

Let $V \in \mathcal{C}$ be transmitted
over the operator channel described in Section~\ref{sec:setup} and assume 
that an $(\ell - \rho + t)$-dimensional subspace $U$ of $W$ is received.
Decoding comprises the following steps:
\begin{enumerate}
\item Invoke {\tt Interpolate}$(U)$ to find a bivariate linearized 
polynomial $Q(x,y)=Q_x(x)+Q_y(y)$ of minimal $(1,k-1)$ weighted degree
that vanishes on the vector space $U$.
\item Invoke {\tt RDiv}$(-Q_x(x),Q_y(x))$ to find a linearized
  polynomial $f(x)$ with the property that $-Q_x(x)\equiv Q_y(x)\otimes f(x)$.
  If no such polynomial can be found declare ``failure.''
\item Output $f(x)$ as the information polynomial corresponding
the codeword $\hat{V} \in \mathcal{C}$ if $d(U,\hat{V})<\ell-k+1$.
\end{enumerate}
The time-complexity of this procedure is dominated by the {\tt Interpolate}
step, which requires $\mathcal{O}((\ell + m)^2)$ operations
in $\F_{q^m}$.

\section{Conclusions}

In this paper we have defined a class of operator channels as the
natural transmission models in ``noncoherent'' random linear
network coding.  The inputs and outputs of operator channels
are subspaces of some given ambient vector space.
We have defined a coding metric on these subspaces which
gives rise to notions of erasures (dimension reduction)
and errors (dimension enlargement).  In defining codes,
it is natural to consider constant-dimension codes;
in this case, the code forms a subset
of a finite-field Grassmannian.
Sphere-packing and sphere-covering
bounds as well as a Singleton-type bound are obtained
in this context.  Finally, a Reed-Solomon-like code construction
(equivalent to the construction of linear authentication
codes in \cite{WXSN03})
is given, and a Sudan-style ``list-1'' unique
decoding algorithm is described, resulting in codes that are capable of
correcting various combinations of errors and erasures.

\section*{Acknowledgments}
The authors wish to thank Ian F. Blake and especially
Danilo Silva for useful comments on an earlier version of this paper.

\bibliographystyle{ieeetr}
\bibliography{refs}

\end{document}